\documentclass[conference,letterpaper]{IEEEtran}
\usepackage[utf8]{inputenc} 
\usepackage[T1]{fontenc}
\usepackage{url}
\usepackage{ifthen}
\usepackage{cite}
\usepackage[cmex10]{amsmath} 
\usepackage[utf8]{inputenc} 
\usepackage[T1]{fontenc}
\usepackage{url}
\usepackage{ifthen}
\usepackage{cite}
\usepackage[cmex10]{amsmath} 
\usepackage{epsfig,epsf,rotating,setspace,latexsym,amssymb,amsfonts,bm,theorem,epstopdf,authblk,bbm,mathrsfs,graphicx,caption, tabularx,dsfont}
\usepackage{subcaption}
\usepackage{algorithm}
\usepackage[noend]{algpseudocode}
\usepackage{color}
\usepackage{mathtools}
\usepackage{multirow}
\usepackage{soul}

\algrenewcommand\algorithmicforall{\textbf{foreach}}
\algrenewcommand\algorithmicindent{.8em}

\newtheorem{theorem}{Theorem}

\newtheorem{remark}{Remark}

\newtheorem{example}{Example}

\title{Private Information Retrieval With Arbitrary Privacy Requirements for Graph-Based Storage}

\author{Mohamed Nomeir \qquad Shreya Meel\qquad  Sennur Ulukus\\
\normalsize Department of Electrical and Computer Engineering\\
\normalsize University of Maryland, College Park, MD 20742 \\
\normalsize \it mnomeir@umd.edu \qquad smeel@umd.edu  \qquad ulukus@umd.edu}
 
\begin{document}

\maketitle

\begin{abstract}
We reformulate the definition of privacy in the private information retrieval (PIR) problem to accommodate flexible privacy requirements. We focus on graph-replicated PIR, with a generalized privacy requirement, instead of requiring all messages to be private from all servers, during retrieval. Towards this, we define a \emph{privacy requirement set} for each server, which can be an arbitrary subset of all message indices, as long as the stored message indices are in their privacy requirement set. Since both the storage and privacy requirement sets have many possibilities, we focus on two specific storage settings, namely the path and cyclic graphs. We consider several privacy settings for each of them, which are not necessarily the same, to give different examples for privacy sets. Of particular interest are the privacy sets that comprise the indices of messages stored at servers within a \emph{neighborhood range}. The neighborhood range parameter allows a transition from the recently introduced local PIR \cite{local_pir} to the standard graph-replicated PIR. In these cases, we derive bounds on the capacity or find the exact capacity.  
\end{abstract}

\section{Introduction}

The original private information retrieval (PIR) problem considers the scenario where a user wishes to retrieve one message out of $K$ messages, while keeping the desired message index private from the servers storing the data. In the original setting, it is assumed that all $K$ messages are fully-replicated across $N$ servers, i.e., $\mathcal{W}_n = \{W_1, W_2, \ldots, W_K\}$, $n \in [N]$, where $\mathcal{W}_n$ is the messages stored at the $n$th server. Several variants under full-replication have been studied extensively in the literature with different threat models \cite{c_pir, c_spir, semantic_pir, banawan_pir_mdscoded, banawan_multimessage_pir, banawan_eaves, nomeir_asymp_bspir, asymmetric,semantic_tpir}.

However, due to the increasing abundance of data, it has become unrealistic to require full message replications across all servers. This motivates a more general setting with arbitrary storage, where each server stores only a subset of the messages. The non fully-replicated PIR model was introduced in \cite{graphbased_pir}. In this model, each message can be replicated any number of times, and stored at $m$ servers, with $m$ ranging from $2$ to $N$, i.e., each message must be stored in at least two servers. It was shown that, this setting can be equivalently modeled using a hypergraph, where the nodes represent the servers and the hyper-edges represent the message indices. For instance, the nodes $i_1, \ldots, i_m\in [N]$ admit a hyper-edge, if the same message is stored at these nodes. Hence, this is known in the literature as (hyper)graph PIR (GPIR). In the following, we denote the capacity for a given graph as $C_{PIR}(\cdot)$. Several variants of this setting can be found in \cite{karim_graph, asymp_gxstpir, star_graph, shreya_graph_4, shreya_graph_1}. 

Due to heterogeneity in data storage, it is impractical to require data privacy equally against all servers. As a simple example, consider a GPIR system with $K_1+K_2$ messages and $N_1+N_2$ servers, that is a disjoint union of two hypergraphs. In particular, let $N_1$ of them follow a certain replication pattern over the $K_1$ messages $W_1, \ldots, W_{K_1}$, while the remaining $N_2$ servers follow another replication pattern over $W_{K_1+1}, \ldots, W_{K_1+K_2}$. Let the required message be $W_1$. To preserve privacy in GPIR, one should download information from the $N_2$ servers, in addition to the $N_1$ servers, even though the information is not useful towards the retrieval of $W_1$, even in removing interference. This motivates the requirement of a privacy definition that is individual for each message, and possibly heterogeneous across servers. As a vital step forward to address this, \emph{local PIR} was introduced recently in \cite{local_pir}. In local PIR, the privacy at each server concerns only with its individual message set, i.e., if a message is not stored by the server, the server may know that the user is retrieving it. It was shown that, the rates and derived capacities are indeed better than their GPIR counterpart. The capacity of local PIR is henceforth denoted by $C_{LPIR}(\cdot)$. 

However, the privacy demands may not be limited to the servers' storage, and could be arbitrary in the real-world. For instance, the retrieval of some sensitive data by the user may need to remain private across all servers, irrespective of whether they store it or not. Towards this, we remove the locality assumption and generalize the privacy definition to encompass more real-world privacy demands. In this setting, we focus on \emph{arbitrary privacy patterns}, i.e., we define a \emph{privacy requirement set} $\mathcal{P}_n$, such that if any two indices $\theta_1, \theta_2 \in \mathcal{P}_n$, then the $n$th server must not know which of the messages $W_{\theta_1}$ or $W_{\theta_2}$ is being retrieved even if $W_{\theta_2} \notin \mathcal{W}_n$. We coin this as PIR with arbitrary privacy requirement sets, which can be viewed as a seamless transition between local PIR and GPIR, by adding indices to each privacy requirement set, i.e., increasing the privacy against each server from its stored message indices to all messages. 

We focus on the path and cyclic graphs, i.e., $\mathbf{P}_N$ and $\mathbf{C}_N$, respectively. For the path graph, we derive the \emph{exact capacity} for the modified-edge-servers privacy setting (see Example~\ref{ex_path}). Next, we consider one and two-sided variants of $h$-hop neighborhood privacy: in the one-sided case, $\mathcal{P}_n$ consists of the indices of all messages present in its own storage and the storage of $h$ servers next to it on one side, while in the two-sided case, $\mathcal{P}_n$ includes the indices of the messages in the storage of $h$ servers in the other side as well. For the path graph, we derive lower bounds on the capacity for the one-sided $h$-neighbor privacy setting, and the two-sided $h$-neighbor privacy setting, with and without modified-edge-server privacy. For the cyclic graph, we find lower and upper bounds on the capacity for the first neighbor privacy setting (see Example~\ref{ex_cycle}) and the $i$th, $i \neq 1$ neighbor privacy setting; the two settings yield completely separate bounds. We also derive a lower bound on the capacity for the one-sided $h$-neighbor privacy setting. Finally, for the cyclic graph, we derive the \emph{exact capacity}  for the two-sided $h$-neighbor privacy setting.

\section{Problem Formulation}
We consider a  system that consists of $N$ servers and $K$ messages, where
the $i$th message has $L_i$ symbols generated uniformly and independently at random, i.e.,
\begin{align}
    H(W_{[K]}) = \sum_{i=1}^K H(W_i) = \sum_{i=1}^K L_i,
\end{align}
where $W_i$ represents the $i$th message, and $[K]$ represents the set $\{1,\ldots,K\}$. As in the case of the usual PIR, the user does not know any side information about the message contents, thus the messages are independent of the queries sent, i.e., 
\begin{align}\label{2}
    I(W_{[K]};\mathcal{Q}) = 0,
\end{align}
where $\mathcal{Q} = \{Q_n^{[k]}:~ n \in [N],~ k \in [K]\}$ and $Q_n^{[k]}$ is the query sent to the $n$th server when retrieving the $k$th message, including the potential null query $Q_n^{[k]} = \emptyset$. The user transmits the query $Q_n^{[k]}$ such that the required message index remains hidden from the intended servers, i.e.,
\begin{align}
    I(\theta; Q_n^{[\theta]}| ~ \theta \in \mathcal{P}_n) = 0, \quad n \in [N], 
\end{align}
where $\mathcal{I}_n \subseteq \mathcal{P}_n \subseteq [K]$ is the set containing the indices of the message that must remain hidden from the $n$th server if being retrieved and $\mathcal{I}_n$ is the set of indices for the messages in the $n$th server, i.e., $\mathcal{I}_n = \{k \in [K]:W_k \in \mathcal{W}_n\}$, where $\mathcal{W}_n$ is the set of messages stored at the $n$th server. It is also important to note that if a server is not queried, it does not know that the scheme is initiated. However, since the scheme is globally known, when a query is transmitted to it, it knows that the previously mentioned message is not being queried from it. The answers generated from the servers are functions of their individual storage, i.e., $\mathcal{W}_n$ and their received query,
\begin{align}\label{4}
    H(A_n^{[k]}| ~ \mathcal{W}_n, ~Q_n^{[k]})=0. 
\end{align}
Once the user receives the answers from all servers, the required message must be decodable, i.e.,
\begin{align}
    \label{decodability}H(W_{\theta}|A_{[N]}^{[\theta]},\mathcal{Q}) = 0.
\end{align}

The rate, i.e., the efficiency, of the scheme is defined as previously defined in the semantic PIR setting \cite{local_pir} since the messages can be of unequal length and the downloaded number of symbols for each message can be different, i.e.,
\begin{align}
    R_{\Pi} = \frac{\mathbb
    {E}\left[L\right]}{\mathbb
    {E}\left[D\right]}=\frac{\sum_{i=1}^K L_i}{\sum_{i=1}^K D_i} = \frac{KL}{\sum_{i=1}^K D_i},
\end{align}
where $\Pi$ is a scheme satisfying \eqref{2}-\eqref{decodability}. Since all messages are equally likely, the rate reduces to the second equality,
where $L_i$ and $D_i = \sum_{n=1}^N H(A_n^{[i]})$ are the $i$th message length and number of downloaded symbols when retrieving the $i$th message, respectively. Although message lengths can be potentially different from each other for different graphs and different privacy sets, we focus here on the setting where the message lengths are equal, $L=L_i$ for all $i$. The capacity $C$ is the supremum over all such $\Pi$.

\section{Motivating Examples}
\begin{example}[Cyclic graph with $1$st neighbor privacy]\label{ex_cycle}
    Let $N=5$, then the storage is given as $\mathcal{W}_1 = \{W_1,W_5\}$, $\mathcal{W}_2 =\{W_1,W_2\}$, $\mathcal{W}_3 = \{W_2,W_3\}$, $\mathcal{W}_4 = \{W_3,W_4\}$, $\mathcal{W}_5 = \{W_4,W_5\}$. The privacy sets are given by
    \begin{align}
        \mathcal{P}_1 & = \{1,2,5\}, ~ \mathcal{P}_2  = \{1,2,3\}, ~
        \mathcal{P}_3  = \{2,3,4\}, \nonumber \\
        \mathcal{P}_4 & = \{3,4,5\}, ~ \mathcal{P}_5  = \{1,4,5\}.
    \end{align}
    Table \ref{table_example_cyclic} shows the retrieval scheme for each message index, where DB denotes database, i.e., server. The message lengths are chosen to be $L=2$. Let $W_1=(a_1, a_2)$, $W_2=(b_1,b_2)$, $W_3=(c_1,c_2)$, $W_4=(d_1,d_2)$ and $W_5=(e_1,e_2)$ be message symbols after uniformly permuting by the user. The rate is $R = \frac{2}{5}$, which is greater than the capacity of the cyclic graph, $C_{PIR}(\mathbf{C}_5) =\frac{1}{3}$ \cite{karim_graph}. The rate is lower than the capacity of local PIR, $C_{LPIR}(\mathbf{C}_5) =\frac{1}{2}$ due to the locality property \cite{local_pir}.

    \begin{table}[h]
        \centering
        \begin{tabular}{|c|c|c|c|c|c|}
        \hline
             & DB 1 & DB 2 & DB 3  & DB 4  & DB 5 \\
             \hline
             $\theta = 1$&  $a_1+e_1$& $a_2+b_1$ &$b_1$  &$d_1$  & $d_1+e_1$ \\
             \hline
             $\theta = 2$ &$a_1+e_1$  &$b_1+a_1$  & $b_2+c_1$ & $c_1$ &$e_1$ \\
             \hline
             $\theta = 3$ & $a_1$ & $b_1+a_1$  & $b_1+c_1$ & $c_2+d_1$ & $d_1$\\
             \hline
             $\theta = 4$ & $e_1$ & $b_1$ & $b_1+c_1$ & $c_1+d_1$ & $d_2+e_1$\\
             \hline
             $\theta = 5$ & $a_1+e_1$ & $a_1$ &  $c_1$& $d_1+c_1$ & $d_1+e_2$\\
             \hline
        \end{tabular}
        \caption{Retrieval scheme for Example~\ref{ex_cycle}.}
        \label{table_example_cyclic}
    \end{table}
\end{example}

\begin{example}[Path \!graph \!with \!modified \!edge \!server \!privacy\!\! ]\label{ex_path}
    Let $N=4$ servers, then the storage isi given as $\mathcal{W}_1  = \{W_1\}$, $\mathcal{W}_2  = \{W_1,W_2\}$, $\mathcal{W}_3 = \{W_2,W_3\}$, $\mathcal{W}_4  = \{W_3\}$. The privacy sets are given by
    \begin{align}
        \mathcal{P}_1 & = \{1,2\}, ~ \mathcal{P}_2  = \{1,2\}, ~ \mathcal{P}_3  = \{2,3\}, ~ \mathcal{P}_4 = \{2,3\}.
    \end{align}
    Table \ref{table_exmple_path} shows the retrieval scheme. The message lengths are chosen to be $L=2$ and the rate is $R = \frac{3}{5}$, which is greater than the capacity of the path graph $C_{PIR}(\mathbf{P}_4) = \frac{1}{2}$\cite{our_journal2025}. We show in Theorem \ref{thm:converse_priv_set1_path} that $\frac{3}{5}$ is indeed the capacity of this setting.
 
    \begin{table}[h]
    \centering
    \begin{tabular}{|c|c|c|c|c|}
    \hline
          &  DB1 & DB 2  & DB 3 & DB 4\\
          \hline
          $\theta = 1$& $a_1$ & $a_2+b_1$ & $b_1$& $\emptyset$\\
          \hline
          $\theta = 2$ & $a_1$ & $b_1+a_1$ & $b_2+c_1$& $c_1$\\
          \hline
          $\theta = 3$ & $\emptyset$ & $b_1$ & $b_1+c_1$ & $c_2$ \\
          \hline
     \end{tabular}
     \caption{Retrieval scheme for Example~\ref{ex_path}.}
     \label{table_exmple_path}
    \end{table}
\end{example}

\section{Results for Path Graphs}

\begin{theorem}[Path \!graph \!with \!modified \!edge \!server \!privacy]\label{thm:converse_priv_set1_path}
    For the path graph with $N$ nodes, $\mathbf{P}_N$ with privacy sets as $\mathcal{P}_1  = \mathcal{P}_2$, $\mathcal{P}_N  = \mathcal{P}_{N-1}$, and $\mathcal{P}_n  = \mathcal{I}_n$, $ n \in [2:N-1]$, the capacity is  
    \begin{align}
        C = \frac{N-1}{2N-3}.
    \end{align}
\end{theorem}
\begin{remark}
    In comparison with the GPIR for path storage, we note that for $N\geq 4$, the capacity for the path graph with modified-edge-server privacy is higher than the capacity for the aforementioned setting, i.e., $\frac{N-1}{2N-3} > \frac{2}{N}$ \cite{our_journal2025}.
\end{remark}

\begin{remark}
    We compare with the path graph with local privacy when $N$ is odd, since in this case the capacity is resolved. In this setting, the capacity of local PIR is higher, where $C_{LPIR}(\mathbf{P}_N) = \frac{N-1}{2N-4} > \frac{N-1}{2N-3}=C(\mathbf{P}_N)$, which stems from the modified-edge-server privacy.
\end{remark}

\subsection{Proof of Theorem \ref{thm:converse_priv_set1_path}}
To find the upper bound, we proceed as follows. Given $\mathbf{P}_N$, let the storage of server $n$ be $\mathcal{W}_1 = \{W_1\}$, $\mathcal{W}_n  =  \{W_n, W_{n-1}\},~ n\in [2:N-1]$ and $\mathcal{W}_N = \{W_{N-1}\}$. Then, for $\theta = k$, we have the download cost $D_k$ bounded as:
\begin{align}
    D_k &\geq \sum_{n=1}^N H(A_n^{[k]}|\mathcal{Q})\geq H(A_{[N]}^{[k]}|\mathcal{Q})\\
    &=   H(W_k|A_{[N]}^{[k]},\mathcal{Q}) +H(A_{[N]}^{[k]}|\mathcal{Q})\\   
    &= H(W_k|\mathcal{Q})+  H(A_{[N]}^{[k]}|W_k,\mathcal{Q})\\
    & = L+ I(\mathcal{W}\setminus \{W_k\};A_{[N]}^{[k]}|W_k,\mathcal{Q})\label{interference_eq}.
\end{align}
The proof relies on showing the following inequalities:
\begin{align}
    &I(\mathcal{W}\setminus\{W_k\};A_{[N]}^{[k]}|W_k, \mathcal{Q})\geq \nonumber \\
    &\begin{cases}
        H(A_2^{[2]}|W_1,\mathcal{Q}), & k=1\\
        \frac {L}{2}+H(A_3^{[3]}|W_1,W_2,\mathcal{Q}), & k=2\\
        H(A_k^{[k-1]}|W_k,\mathcal{Q})+H(A_{k+1}^{[k+1]}|\mathcal{W}_{k},\mathcal{Q}), & k\in [3:N-3]\\
        \frac{L}{2}+H(A_{N-2}^{[N-3]}|W_{N-1}, W_{N-2}, \mathcal{Q}), & k=N-2\\
        H(A_{N-1}^{[N-2]}|W_{N-1},\mathcal{Q}), & k=N-1.
    \end{cases}
\end{align}
Note that the second and fourth inequalities introduced in this setting, while the others are the same as those in local PIR \cite{local_pir}. This is because the number of servers that queries for $W_2$ and $W_{N-2}$ require privacy from, has increased, compared to that in local PIR. We show the proof of $k=2$, and the case of $k=N-2$ follows similarly,
\begin{align}
    I&(\mathcal{W}\setminus \{W_2\};A_{[N]}^{[2]}|W_2,\mathcal{Q}) \nonumber \\ &\geq I(W_1,W_3;A_1^{[2]},A_2^{[2]},A_3^{[2]}|W_2, \mathcal{Q})\\
    &\geq I(W_1;A_1^{[2]},A_2^{[2]}|W_2, \mathcal{Q}) + I(W_3;A_3^{[2]}|W_1,W_2,\mathcal{Q})\\
    &\geq \frac{1}{2}I(W_1;A_1^{[2]}|W_2,\mathcal{Q})+\frac{1}{2}I(W_1;A_2^{[2]}|W_2,\mathcal{Q}) \nonumber \\ & \quad+  I(W_3;A_3^{[2]}|W_1,W_2,\mathcal{Q}) \label{c_1}\\
    & = \frac{1}{2}H(A_1^{[1]}|W_2,\mathcal{Q})+\frac{1}{2}H(A_2^{[1]}|W_2,\mathcal{Q})\nonumber \\ & \quad +H(A_3^{[3]}|W_1,W_2,\mathcal{Q})\label{c_2}\\
    &\geq \frac{1}{2}H(A_1^{[1]},A_2^{[1]}|W_2,\mathcal{Q}) +\frac{1}{2}H(W_1|A_1^{[1]},A_2^{[1]},W_2,\mathcal{Q}) \notag\\ &\quad+H(A_3^{[3]}|W_1,W_2,\mathcal{Q})\\
    &= \frac{L}{2} + H(A_3^{[3]}|W_1,W_2,\mathcal{Q}),
\end{align}
where $\eqref{c_1}$ is due to $2 I(W_1;A_1^{[2]},A_2^{[2]}|W_2,\mathcal{Q}) = I(W_1;A_1^{[2]}|W_2,\mathcal{Q})+ I(W_1;A_2^{[2]}|A_1^{[2]}, W_2,\mathcal{Q})+ I(W_1;A_2^{[2]}|W_2,\mathcal{Q})+ I(W_1;A_1^{[2]}|A_2^{[2]}, W_2,\mathcal{Q})$, and \eqref{c_2} is due to privacy. Now, by summing over all $k$, we obtain,
\begin{align}
    & \sum_{k=1}^{N-1}  D_k -(N-1)L \nonumber \\ &\geq  H(A_2^{[2]}|W_1,\mathcal{Q})+ \frac{L}{2} +H(A_3^{[3]}|W_1,W_2,\mathcal{Q})\notag \\
    &\quad +\sum_{k=3}^{N-3} \left( H(A_k^{[k-1]}|W_k,\mathcal{Q}) +H(A_{k+1}^{[k+1]}|W_k,  W_{k-1},\mathcal{Q}) \right)\notag\\
    &\quad+ \frac{L}{2}+H(A_{N-2}^{[N-3]}|W_{N-1},W_{N-2},\mathcal{Q}) \nonumber \\ &\quad+H(A_{N-1}^{[N-2]}|W_{N-1},\mathcal{Q})\\
    &= L+ H(A_2^{[2]}|W_1,\mathcal{Q})+ H(A_3^{[3]}|W_1,W_2,\mathcal{Q})\notag\\
    &\quad+ H(A_3^{[2]}|W_3, \mathcal{Q}) + H(A_4^{[4]}|W_3,W_2,\mathcal{Q})\notag \\
    &\quad  +\ldots+ H(A_{N-3}^{[N-4]}|W_{N-3},\mathcal{Q}) \nonumber \\ &\quad+H(A_{N-2}^{[N-2]}|W_{N-2},W_{N-3}, \mathcal{Q}) \notag\\
    &\quad+H(A_{N-1}^{[N-2]}|W_{N-1},\mathcal{Q})\\
    &\geq L+ H(A_2^{[2]},A_3^{[2]}|\mathcal{W}\setminus \{W_2\},\mathcal{Q})\nonumber\\ &\quad+\sum_{k=3}^{N-2} H(A_k^{[k]}, A_{k+1}^{[k]}|\mathcal{W}\setminus \{W_k\},\mathcal{Q})\\
    &= L+ H(W_2,A_2^{[2]},A_3^{[2]}|\mathcal{W}\setminus \{W_2\},\mathcal{Q})\nonumber \\ & \quad +\sum_{k=3}^{N-2} H(W_k,A_k^{[k]}, A_{k+1}^{[k]}|\mathcal{W}\setminus \{W_k\},\mathcal{Q})\\
    &= L+H(W_2|\mathcal{W}\setminus \{W_2\},\mathcal{Q})+H(A_2^{[2]},A_3^{[2]}|\mathcal{W},\mathcal{Q})\notag \\
    &\quad+\left(  \sum_{k=3}^{N-2} H(W_k|\mathcal{W}\setminus\{W_k\},\mathcal{Q})+ H(A_k^{[k]}, A_{k+1}^{[k]}|\mathcal{W},\mathcal{Q})\right)\\
    &= (N-2)L.
\end{align}

For achievability, we have two cases. The first case is $\theta \in \{1, N-1\}$ and the second case is $\theta \in [2:N-2]$. For each $k\in [N-1]$, let $W_k(i)$ denote the $i$th symbol of $W_k$ after permuting the indices of the messages uniformly at random. In the first case, we focus on $\theta =1$ and the other case is handled similarly. In this case, we download $W_1(1)$ from server $1$, $W_1(2)+W_2(1)$ from server $2$ and $W_2(1)$ from server $3$. As for the second case, assume without loss of generality that $\theta =2$, then we download $W_1(1)$ from server $1$, $W_1(1)+W_2(1)$ from server $2$ and $W_2(2)+W_3(1)$ from server $3$ and $W_3(1)$ from server $4$. The total downloaded symbols is given by $D = 2(3) + (N-3)(4) = 2(2N-3)$ and the rate is given by $R = \frac{2(N-1)}{2(2N-3)} = \frac{N-1}{2N-3}$, concluding the achievability proof.

The next theorem provides a capacity lower bound for path graphs as a function of $h$.
\begin{theorem}[Path \!graph \!with \!one-sided \!$h$-neighbor \!privacy]
    \label{thm_h_shift_one-sided_path}
     For $\mathbf{P}_N$, with privacy sets 
     \begin{align}
         \mathcal{P}_n = [\max(1,n-1):\min(n+h,N-1)],
     \end{align}
     the capacity is lower-bounded by 
    \begin{align}
        C \geq \frac{2(N-1)}{\frac{(h+2)(h+1)}{2}+3h+5+(h+4)(N-h-3)},
    \end{align}
   where $h\in \{1,\ldots,N-2\}$.
\end{theorem}

\begin{remark}\label{rmk_c_lpir lower}
    When $h=0$, this reduces to the local privacy setting. In this case, we use the scheme for local PIR in \cite{local_pir} where $C_{LPIR}(\mathbf{P}_N)$ is lower bounded by $\frac{N-1}{2N-3}$ when $N$ is even and is equal to $\frac{N-1}{2N-4}$ when $N$ is odd \cite{local_pir}. Interestingly, since there is no value of $h$ for which the privacy sets cover all message indices, we do not recover the GPIR setting.
\end{remark}
\begin{remark}
    The main reason why the case $h = 0$ is not considered here is because the scheme for $h=0$ is dependent on the bipartite graph structure of $\mathbf{P}_N$ and requires downloading all messages in $\mathcal{W}_n$ whenever server $n$ is contacted, for privacy. However, for $h>0$, as the privacy sets become larger and non-symmetric, the bipartite scheme becomes less efficient.
\end{remark}

\subsection{Proof of Theorem \ref{thm_h_shift_one-sided_path}}
There are three cases in this setting. The first is when $\theta \in [1:h+1]$, where we download $\theta+2$ symbols from servers $1$ to server $\theta+2$ as follows. If $\theta =1$, we download $W_{1}(1)$ from server $1$ and use it as an information symbol. From the second server, we download $W_1(2)+W_2(1)$ and from the third server, we download $W_{2}(1)$. If $\theta\neq 1$, we download $W_1(1)$ to be used as side information. Further, we download $W_{\theta - 1}(1)+W_{\theta}(1)$ from the $\theta$th server and $W_{\theta}(2)+W_{\theta + 1}(1)$ from the $\theta+1$th server, $W_{\theta + 1}(1)$ from $\theta+2$th server and $W_{k-1}(1)+W_k(1)$ from the $k$th server, where $k \in [2:\theta-1]$. The second case is when $\theta \in [h+2:N-2]$; we download $h+4$ symbols starting with server $k-1$ when $\theta = h+k $ as a clean symbol for the largest index message, i.e., $W_{k}(1)$, which acts as side information. Similarly, server $k+h+2$ acts as a source of side information, where $W_{k+h+1}(1)$ is downloaded, and for the rest of the servers in between, we download $W_{m-1}(1)+W_{m}(1)$ for $m \in [k:k+h+1] \setminus \{\theta-1,\theta\}$, $W_{\theta-1}(1)+W_{\theta}(1)$ from the $\theta$th server and $W_{\theta}(2)+W_{\theta+1}(1)$ from the $\theta+1$th server. As for the last case when $\theta = N-1$, we download one symbol from the last $h+3$ servers in a similar fashion. Thus, the sum of all downloaded symbols is given by
\begin{align}
    \sum_{\theta = 1}^{N-1} D_{\theta} &= \left(\sum_{\theta=1}^{h+1} \theta+2\right) + \left(\sum_{\theta = h+2}^{N-2} h+4\right) + h+3 \\
    &= \frac{(h+2)(h+1)}{2}+3h+5+(h+4)(N-h-3),
\end{align}
and the rate follows.

\begin{theorem}\label{thm:h_shift two-sided_rate path}
    Consider the privacy sets for server $n$ in $\mathbf{P}_N$ given by either
    \begin{enumerate}
        \item Two-sided $h$-neighbor privacy: 
        \begin{align}
            \mathcal{P}_n = [\max(1,n-h-1): \min(n+h,N-1)],
        \end{align}
        where $h\in \{1,\ldots,N-3\}$.
        \item Two-sided $h$-neighbor privacy with modified edge servers privacy: $\mathcal{P}_1 = \mathcal{P}_2$, $\mathcal{P}_{N-1} = \mathcal{P}_N$, and for $n\in [2:N-1]$
        \begin{align}
           \mathcal{P}_n &= [\max(1,n-h-1): \min(n+h,N-1)],        
        \end{align}
        where $h\in \{0,1,\ldots,N-3\}$.
    \end{enumerate}
    In both cases, the capacity is lower-bounded as,
    \begin{align}
        C \geq \frac{2(N-1)}{(h+2)(2N-h-3)}.
    \end{align}
\end{theorem}
The proof of Theorem~\ref{thm:h_shift two-sided_rate path} can be found in Appendix \ref{proof_of_thm_3}.

\begin{remark}
When $h=0$, for the first case, we recover the local privacy setting, and we obtain the same result as Remark~\ref{rmk_c_lpir lower}. For the second case, we recover the modified-edge-server privacy setting, as in Theorem~\ref{thm:converse_priv_set1_path}. On the other hand, when $h=N-3$, each privacy set is $[K]$, yielding $C_{PIR}(\mathbf{P}_N)=\frac{2}{N}$.    
\end{remark}

\section{Results for Cyclic Graphs}
To make the presentation concise, we note that all additions and subtractions are performed modulo $N$, hereafter and $N$ is used instead of $0$, i.e., $W_0 = W_N$. Moreover, when $a> b$, $[a:b]$ denotes the set $\{a,a+1, \ldots, N, 1, 2, \ldots,b\}$.

\begin{theorem}[Cyclic graph with $1$st neighbor privacy]\label{thm:cyclic-1st}
    For the cyclic graph with $N$ nodes, $\mathbf{C}_N$, with $\mathcal{P}_n = \mathcal{I}_n \cup \mathcal{I}_{n+1} $, the capacity is lower bounded by,
    \begin{align}
       \frac{2}{5} \leq C \leq \frac{1}{2}.
    \end{align}
\end{theorem}

\begin{remark}
  Note that, both lower and upper bounds are independent of the number of servers $N$. This is because the privacy for each index is confined to a fixed number of servers, due to the symmetry of cyclic graphs, unlike the path graphs. 
\end{remark}

\begin{remark}
    For $N \geq 5$, the lower bound on the capacity with the $1$st neighbor privacy is higher than the capacity of GPIR for the cyclic graph, where $C_{PIR}(\mathbf{C}_N) = \frac{2}{N+1}$ \cite{karim_graph}.
\end{remark}

\subsection{Proof of Theorem \ref{thm:cyclic-1st}}
For the lower bound, we proceed as follows. By the symmetry of $\mathbf{C}_N$, assume that $\theta=1$ without loss of generality. Let $L=2$ symbols for all messages. The user permutes the message indices using permutations chosen uniformly at random. The user downloads $W_1(1)+W_N(1)$ from server $1$ and $W_1(2)+W_2(1)$ from server $2$. To decode the two message symbols, the user downloads $W_{2}(1)$ from server $3$, $W_N(1)+W_{N-1}(1)$ from server $N$ and $W_{N-1}(1)$ from server $N-1$. Thus, $5$ symbols are downloaded to decode the required two message symbols.

For the upper bound, we have $C_{LPIR}(\mathbf{C}_N)=\frac{1}{2}$. Since the local PIR capacity is always higher than the arbitrary private set capacity, we have the upper bound.

\begin{theorem}[Cyclic graph with $i$th neighbor privacy]\label{thm:cyclic_k_k+i}
    For the cyclic graph with $N$ nodes, $\mathbf{C}_N$, and $\mathcal{P}_n = \mathcal{I}_n \cup \mathcal{I}_{n+i} $, $i \in [2:N-2]$, the capacity is bounded by,
    \begin{align}
       \frac{1}{3} \leq C \leq \frac{2}{5} .
    \end{align}
\end{theorem}
The proof of Theorem~\ref{thm:cyclic_k_k+i} can be found in Appendix \ref{pf_thm_5}.

\begin{remark}
    We have $\frac{2}{5}$ here as an upper bound as opposed to a lower bound for Theorem \ref{thm:cyclic-1st}. This is because the number of servers that have $k \in \mathcal{P}_n$ has increased for all $k \in [K]$, requiring stricter privacy.
\end{remark}

\begin{theorem}[Cyclic \!graph \!with \!one-sided \!$h$-neighbor\! privacy\!]\label{thm:h_shift one-sided_rate cycle}
    Consider the privacy sets for server $n$ in $\mathbf{C}_N$ given by,
    \begin{align}
    \mathcal{P}_n = [n-1:n+h],
    \end{align}
    where $h\in \{\!0,\!1,\!\ldots\!,N-2\!\}$. The capacity is lower-bounded as
    \begin{align}
        C \geq \frac{2}{h+4}.
    \end{align}
\end{theorem}
The proof of Theorem~\ref{thm:h_shift one-sided_rate cycle} can be found in Appendix \ref{pf_thm_6}.

\begin{remark}
    The lower bound coincides with the PIR capacity, $\frac{2}{N+1}$ when $h=N-3$, whereas $\mathcal{P}_n = [N]$, $n\in [N]$ when $h=N-2$. This gap suggests the existence of another scheme with rate $R = \frac{2}{h+3}$, which might be the capacity of this setting.
\end{remark}

\begin{theorem}[Cyclic \!graph \!with \!two-sided \!$h$-neighbor\! privacy\!]\label{thm:h_shift two-sided_rate cycle}
    Consider the privacy sets for server $n$ in $\mathbf{C}_N$ given by,
    \begin{align}
    \mathcal{P}_n = [n-h-1:n+h],
    \end{align}
    where $h\in \{0,1,\ldots,\lfloor\frac{N-3}{2}\rfloor\}$. The capacity is 
    \begin{align}
        C = \frac{1}{h+2}.
    \end{align}
\end{theorem}

\begin{remark}
    When $h=0$, the capacity results coincide with that for local PIR, i.e., $C_{LPIR}(\mathbf{C}_N) = \frac{1}{2}$ \cite{local_pir}.
\end{remark}

\subsection{Proof of Theorem \ref{thm:h_shift two-sided_rate cycle}}
To prove the achievability, let $\theta = 1$. Then, the index needs to remain private from servers $3$ to $3+h-1$, i.e., $h$ servers starting from server $2$. Similarly, it needs to be private from servers $N$ to $N-h+1$. From server $1$, $W_1(1)+W_N(1)$, and from serve 2, $W_1(2)+W_2(1)$ is downloaded. From server $k \in [3:3+h-1]$, $W_{k-1}(1)+W_k(1)$ is downloaded, similarly, for $k \in [N-h+1:N]$. Then, for server $3+h$, $W_{3+h-1}(1)$ is download and the same goes for server $N-h$. Thus, the rate is $R = \frac{2N}{2(h+2)N} = \frac{1}{h+2}$. To prove the converse, we have, as \eqref{interference_eq}, for any $k\in [N]$, that $D_k \geq L + I(\mathcal{W}\setminus \{W_k\};A_{[N]}^{[k]}|W_k,\mathcal{Q})$. Let $\mathcal{W}'' \!\!=\!\! \{\!W_{k-h-1}, \ldots,W_{k-1},W_{k+1},\ldots,W_{k+h+1}\!\}$, then
\begin{align}
    I(\mathcal{W}&\setminus \{W_k\};A_{[N]}^{[k]}|W_k,\mathcal{Q})\geq I(\mathcal{W}'';A_{[N]}^{[k]}|W_k,\mathcal{Q})\\
    =& \sum_{l=k-h}^{k} I(W_{l-1};A_{[N]}^{[k]}|W_{[l:k]},\mathcal{Q}) \nonumber \\ 
    &\quad +\sum_{l=k+1}^{k+h+1} I(W_{l};A_{[N]}^{[k]}|W_{[k-h-1:l-1]},\mathcal{Q})\\
    \geq& \sum_{l=k-h}^{k} I(W_{l-1};A_l^{[k]}|\mathcal{W}\setminus \{W_{l-1}\},\mathcal{Q})\nonumber \\ 
    & \quad +\sum_{l=k+1}^{k+h+1} I(W_{l};A_l^{[k]}|\mathcal{W}\setminus \{W_l\},\mathcal{Q})\\
    =& \sum_{l=k-h}^{k} I(W_{l-1};A_l^{[l-1]}|\mathcal{W}\setminus \{W_{l-1}\},\mathcal{Q})\nonumber \\ 
    & \quad +\sum_{l=k+1}^{k+h+1} I(W_{l};A_l^{[l]}|\mathcal{W}\setminus \{W_l\},\mathcal{Q})\\
    =& \sum_{l=k-h}^k H(A_l^{[l-1]}|\mathcal{W}\setminus \{W_{l-1}\},\mathcal{Q}) \nonumber \\ 
    & \quad +\sum_{l=k+1}^{k+h+1} H(A_l^{[l]}|\mathcal{W}\setminus \{W_l\},\mathcal{Q}).
\end{align}
Then, by summing over all $k \in [N]$,
\begin{align}
    \sum_{k=1}^N D_k-NL &\geq \sum_{k=1}^N\sum_{l=k-h}^k H(A_l^{[l-1]}|\mathcal{W}\setminus \{W_{l-1}\},\mathcal{Q}) \nonumber \\ & \quad +\sum_{k=1}^N \sum_{l=k+1}^{k+h+1} H(A_l^{[l]}|\mathcal{W}\setminus \{W_l\},\mathcal{Q})\\\
    &=(h+1)\sum_{l=1}^N H(A_l^{[l]}|\mathcal{W}\setminus \{W_l\},\mathcal{Q}) \nonumber \\ & \quad +(h+1)\sum_{l=1}^NH(A_{l+1}^{[l]} |\mathcal{W}\setminus \{W_{l}\},\mathcal{Q})\\
    &\geq \!(h+1) \!\sum_{l=1}^N \!H(\!A_l^{[l]},\! A_{l+1}^{[l]}|\mathcal{W}\setminus \{W_l\},\!\mathcal{Q})\\
    &= (h+1)NL.
\end{align}

\bibliographystyle{unsrt}
\bibliography{references_2.bib}
\clearpage

\appendix

\subsection{Proof of Theorem~\ref{thm:h_shift two-sided_rate path}}\label{proof_of_thm_3}
For better exposition, we explain the schemes with reference to the GPIR scheme on path graph given in \cite{our_journal2025}. First, we demonstrate the scheme for 1). Assume $L_i=2$ for all $i\in [K]$. Let $\theta=k$. The set of servers $\mathcal{N}_k$ that have $k$ in their privacy set is given by
\begin{align}
    \mathcal{N}_k = [\max(1,k-h):\min(N,h+k+1)].
\end{align}
We analyze this in two cases:
\paragraph*{Case 1: $h\leq \lceil\frac{N-4}{2}\rceil$} In this range, no message index appears in all $\mathcal{P}_n, n\in [N]$. This results in the following three ranges of $k$:
\begin{enumerate}
    \item If $k\in [h+1]$, then $\mathcal{N}_k = [1:h+k+1]$. Download one symbol corresponding to the PIR scheme on path graph consisting of $\mathcal{N}_k$ and server $h+k+2$, when $\theta=k$.
    \item If $k\in [h+2:N-h-2]$, then $\mathcal{N}_k = [k-h:h+k+1]$. Download one symbol corresponding to the PIR scheme on path graph consisting of $\mathcal{N}_k$, server $k-h-1$ and server $h+k+2$, when $\theta=k$.
    \item If $k\in [N-h-1:N-1]$, then $\mathcal{N}_k = [k-h:N]$. Download one symbol corresponding to the PIR scheme on path graph consisting of $\mathcal{N}_k$ and server $k-h-1$, when $\theta=k$.
\end{enumerate}
Let $M_k = |\mathcal{N}_k|+1$, $c_{1,k}=h+k+2$, $c_{2,k} = 2h+4$ and $c_{3,k} = N-k+h+2$, then,
\begin{align}
    R&=\frac{2(N-1)}{\sum_{k=1}^{h+1}M_k+\sum_{k=h+2}^{N-h-2}(M_k+1)+\sum_{k=N-h-1}^{N-1}M_k}\\
    &=\frac{2(N-1)}{\sum_{k=1}^{h+1}c_{1,k}+\sum_{k=h+2}^{N-h-2}c_{2,k}+\sum_{k=N-h-1}^{N-1} c_{3,k}}\\
    &=\frac{N-1}{\sum_{k=1}^{h+1}(h+k+2)+(h+2)(N-2h-3)}\\
    &=\frac{2(N-1)}{(h+2)(2N-h-3)}.
\end{align}

\paragraph*{Case 2: $h>\lceil\frac{N-4}{2} \rceil$} In this range, there exists at least one $k$ that is present in the privacy set of all servers. This results in the following three ranges of $k$:
\begin{enumerate}
    \item If $k\in [1:N-h-2]$, then $\mathcal{N}_k = [1:h+k+1]$. Download one symbol corresponding to the PIR scheme on path graph consisting of $\mathcal{N}_k$ and server $h+k+2$, when $\theta=k$.
    \item If $k\in [N-h-1:h+1]$, then $\mathcal{N}_k = [N]$. Download one symbol corresponding to the PIR scheme on path graph $\mathbf{P}_N$, when $\theta=k$.
    \item If $k\in [h+2:N-1]$, then $\mathcal{N}_k = [k-h:N]$. Download one symbol corresponding to the PIR scheme on path graph consisting of $\mathcal{N}_k$ and server $k-h-1$, when $\theta=k$.
\end{enumerate}
\begin{align}
    R&=\frac{2(N-1)}{\sum_{k=1}^{N-h-2}M_k+\sum_{k=N-h-1}^{h+1}N+\sum_{k=h+2}^{N-1}M_k}\\
    &=\frac{2(N-1)}{\sum_{k=1}^{N-h-2}c_{1,k}+\sum_{k=N-h-1}^{h+1}N+\sum_{k=h+2}^{N-1} c_{3,k}}\\
    &=\frac{2(N-1)}{2\sum_{k=1}^{N-h-2}(h+k+2)+(2h+3-N)N}\\
    &=\frac{2(N-1)}{(h+2)(2N-h-3)}.
\end{align}
The rate for 2) can be derived similarly, with the following minor change. The number of servers $\mathcal{N}'_{k}$ that have $k$ in their privacy sets is:
\begin{align}
    \mathcal{N}_k' = 
    \begin{cases}
        \mathcal{N}_{h+2} \cup \{1\}, & k=h+2,\\
        \mathcal{N}_{N-h-2} \cup \{N\}, & k = N-h-2,\\
        \mathcal{N}_k, & \text{ otherwise.}
    \end{cases}
\end{align}
We analyze this in two cases:

\paragraph*{Case 1: $h\leq \lceil\frac{N-6}{2}\rceil$} In this range, no message index appears in all $\mathcal{P}_n, n\in [N]$. This results in the following three ranges of $k$:
\begin{enumerate}
    \item If $k\in [h+2]$, then $\mathcal{N}_k' = [1:h+k+1]$. Download one symbol corresponding to the PIR scheme on path graph consisting of $\mathcal{N}_k'$ and server $h+k+2$, when $\theta=k$.
    \item If $k\in [h+3:N-h-3]$, then $\mathcal{N}_k' = [k-h:h+k+1]$. Download one symbol corresponding to the PIR scheme on path graph consisting of $\mathcal{N}_k'$, server $k-h-1$ and server $h+k+2$, when $\theta=k$.
    \item If $k\in [N-h-2:N-1]$, then $\mathcal{N}_k' = [k-h:N]$. Download one symbol corresponding to the PIR scheme on path graph consisting of $\mathcal{N}_k'$ and server $k-h-1$, when $\theta=k$.
\end{enumerate}
This gives the same rate as shown below, since,
\begin{align}
    R&=\frac{2(N-1)}{2\sum_{k=1}^{h+1}M'_k+\sum_{k=h+3}^{N-h-3}2M'_k+1+\sum_{k=N-h-1}^{N-1}M'_k}\\
    &=\frac{2(N-1)}{\sum_{k=1}^{h+1} M_k+\sum_{k=h+2}^{N-h-2}M_k+1 +\sum_{k=N-h-1}^{N-1}M_k}\\
    &=\frac{2(N-1)}{(h+2)(2N-h-3)},
\end{align}
where $M'_k = |\mathcal{N}'_k|+1$.

\paragraph*{Case 2: $h>\lceil\frac{N-6}{2} \rceil$} In this range, there exists at least one $k$ that is present in the privacy set of all servers. This results in the following three ranges of $k$:
\begin{enumerate}
    \item If $k\in [1:N-h-3]$, then $\mathcal{N}_k' = [1:h+k+1]$. Download one symbol corresponding to the PIR scheme on path graph consisting of $\mathcal{N}_k$ and server $h+k+2$, when $\theta=k$.
    \item If $k\in [N-h-2:h+2]$, then $\mathcal{N}_k' = [N]$. Download one symbol corresponding to the PIR scheme on path graph $\mathbf{P}_N$, when $\theta=k$.
    \item If $k\in [h+3:N-1]$, then $\mathcal{N}_k' = [k-h:N]$. Download one symbol corresponding to the PIR scheme on path graph consisting of $\mathcal{N}_k$ and server $k-h-1$, when $\theta=k$.
\end{enumerate}
\begin{align}
    R&=\frac{2(N-1)}{\sum_{k=1}^{N-h-3}M'_k+\sum_{k=N-h-2}^{h+2}N+\sum_{k=h+3}^{N-1}M'_k}\\
    &=\frac{2(N-1)}{\sum_{k=1}^{N-h-3}c_{1,k}+\sum_{k=N-h-2}^{h+2}N+\sum_{k=h+3}^{N-1} c_{3,k}}\\
    &=\frac{2(N-1)}{(h+2)(2N-h-3)}.
\end{align}

\subsection{Proof of Theorem \ref{thm:cyclic_k_k+i}}\label{pf_thm_5}
To find the lower bound, we proceed as follows. Without loss of generality, let $\theta = 1$. From the second server, we download $W_1(2)+W_2(1)$. From the third server, we always download $W_2(1)$. From server $1$, we download $W_1(1)+W_N(1)$. To cancel the interference here, we have two cases, $i=2$ and $i \neq 2$. If $i=2$, we download $W_N(1)+W_{N-1}(1)$ from server $N$, $W_{N-1}(1)+W_{N-2}(1)$ from server $N-1$ and $W_{N-2}(1)$ from server $N-2$. As for the case of $i \neq 2$, we download $W_{N}(1)$ from server $N$ and download any linear combination from the servers that has $1 \in \mathcal{P}_n$. Thus, the rate is given by $R = \frac{2N}{6N} = \frac{1}{3}$.

As for the upper bound, recall that
\begin{align}
    D_k \geq L + I(\mathcal{W}\setminus W_k; A_{[N]}^{[k]}|W_k,\mathcal{Q}).
\end{align}
Let $\mathcal{W}^* = \{W_k,W_{k-1},W_{k-i+1},W_{k-i+3},W_{k+1}\}$, then,
\begin{align}
    I(&\mathcal{W}\setminus W_k; A_{[N]}^{[k]}|W_k,\mathcal{Q})\nonumber \\ \geq& I(W_{k-i+1},W_{k-1};A_{[N]}^{[k]}|W_k,\mathcal{Q}) \nonumber \\ &+ I(W_{k-i+3},W_{k+1};A_{[N]}^{[k]}|W_k,W_{k-1},W_{k-i+1},\mathcal{Q}) \nonumber \\
    &+ I(W_{k-i-1},W_{k-i};A_{[N]}^{[k]}|\mathcal{W}^*,\mathcal{Q}) \\
    \geq & I(W_{k-i+1},W_{k-1};A_{k}^{[k]}|W_k,\mathcal{Q}) \nonumber \\ &+ I(W_{k-i+3},W_{k+1};A_{k+1}^{[k]}|W_k,W_{k-1},W_{k-i+1},\mathcal{Q}) \nonumber \\
    &+ I(W_{k-i-1},W_{k-i};A_{k-i}^{[k]},A_{k-i+1}^{[k]}|\mathcal{W}^*,\mathcal{Q})\\
    =& H(A_k^{[k-1]}|W_k,\mathcal{Q}) + H(A_{k+1}^{[k+1]}|W_k,W_{k-1},W_{k-i+1,},\mathcal{Q}) \nonumber \\
    &+ I(W_{k-i-1},W_{k-i};A_{k-i}^{[k]},A_{k-i+1}^{[k]}|\mathcal{W}^*,\mathcal{Q}). \label{c_1_1}
\end{align}
Now, we prove that the third term in \eqref{c_1_1} is lower bounded by $\frac{L}{2}$. For convenience, we denote $\mathcal{W}' = \{W_{k-i-1},W_k,W_{k-1},W_{k-i+1},W_{k-i+3},W_{k+1}\}$. Thus,
\begin{align}
     I(&W_{k-i-1},W_{k-i};A_{k-i}^{[k]},A_{k-i+1}^{[k]}|\mathcal{W}' \setminus W_{k-i-1},\mathcal{Q}) \nonumber \\ &\geq I(W_{k-i};A_{k-i}^{[k]},A_{k-i+1}^{[k]}|\mathcal{W}' ,\mathcal{Q})\\
     &\geq \frac{1}{2}I(W_{k-i};A_{k-i}^{[k]}|\mathcal{W}' ,\mathcal{Q})+ \frac{1}{2}I(W_{k-i};A_{k-i+1}^{[k]}|\mathcal{W}' ,\mathcal{Q})\\
     &= \frac{1}{2}I(W_{k-i};A_{k-i}^{[k-i]}|\mathcal{W}' ,\mathcal{Q})+ \frac{1}{2}I(W_{k-i};A_{k-i+1}^{[k-i]}|\mathcal{W}' ,\mathcal{Q})\\
     &= \frac{1}{2}H(A_{k-i}^{[k-i]}|\mathcal{W}' ,\mathcal{Q}) + \frac{1}{2}H(A_{k-i+1}^{[k-i]}|\mathcal{W}' ,\mathcal{Q})\\
     & \geq \frac{1}{2}H(A_{k-i}^{[k-i]}|\mathcal{W}\setminus W_{k-i} ,\mathcal{Q}) \nonumber \\ &\quad + \frac{1}{2}H(A_{k-i+1}^{[k-i]}|\mathcal{W} \setminus W_{k-i} ,\mathcal{Q})\\
     &\geq \frac{1}{2}H(A_{k-i}^{[k-i]},A_{k-i+1}^{[k-i]}|\mathcal{W}\setminus W_{k-i} ,\mathcal{Q})\\
     & = \frac{1}{2}H(W_{k-i},A_{k-i}^{[k-i]},A_{k-i+1}^{[k-i]}|\mathcal{W}\setminus W_{k-i} ,\mathcal{Q})\\
     &=\frac{L}{2}.
\end{align}
Thus, summing over $k \in [N]$, we have
\begin{align}
    \sum_{k=1}^N D_k & \geq \frac{3NL}{2} + \sum_{k=1}^N H(A_k^{[k-1]}|W_k,\mathcal{Q}) \nonumber \\ & \quad+ H(A_{k+1}^{[k+1]}|W_k,W_{k-1},W_{k-i+1,},\mathcal{Q})\\
    & \geq \frac{5NL}{2},
\end{align}
where the last inequality follows the same way as in the proof of Theorem \ref{thm:converse_priv_set1_path}.

\subsection{Proof of Theorem \ref{thm:h_shift one-sided_rate cycle}}\label{pf_thm_6}
Assume $L=2$ is the message length and $\theta=k$. Let the user independently permute the symbols of each message, uniformly at random, and denote the permuted message as $W_k = [W_k(1), W_k(2)]$. The set of servers $\mathcal{N}_k$ that have $k$ in their privacy set is given by
\begin{align}
    \mathcal{N}_k = [k-h:k+1].
\end{align}
From server $\ell \in [k-h:k]$, the user queries the sum $W_{\ell-1}(1)+W_\ell(1)$. From server $k+1$, the user downloads the sum $W_{k}(2)+W_{k+1}(1)$. Now, from server $k-h-1$, the user downloads $W_{k-h-1}(1)$, and from server $k+2$, the user downloads $W_{k+1}(1)$ as interference symbols. These are used to cancel the unwanted message symbols and recover $W_k$. Thus, $(h+2)+2$ symbols are downloaded to recover $2$ symbols of $W_k$, resulting in the rate $R=\frac{2}{h+4}$.

\end{document}